**Title:** An Application of the Causal Roadmap in Two Safety Monitoring Case Studies: Covariate-Adjustment and Outcome Prediction using Electronic Health Record Data

**Authors:**
Brian D Williamson*[1], Richard Wyss*[2], Elizabeth A Stuart[3], Lauren E Dang[4], Andrew N Mertens[4], Andrew Wilson[5], Susan Gruber[6]
*co-first authors

**Author Affiliations:**
[1]Biostatistics Division, Kaiser Permanente Washington Health Research Institute, Seattle, WA, USA
[2]Division of Pharmacoepidemiology and Pharmacoeconomics, Brigham and Women's Hospital, Harvard Medical School
[3]Department of Mental Health, Johns Hopkins Bloomberg School of Public Health, Baltimore, MD, USA
[4]Department of Biostatistics, University of California, Berkeley, CA, USA
[5]Parexel International, Durham, NC, USA
[6]TL Revolution, Cambridge, MA, USA

*Source of Funding:* This work received no specific grant from any funding agency in the public, commercial, or not-for-profit sectors

*Conflicts of interest:*
**LED** reports tuition and stipend support from a philanthropic gift from the Novo Nordisk corporation to the University of California, Berkeley to support the Joint Initiative for Causal Inference. **SG** reports that she is a co-founder of the statistical software start-up company TLrevolution, Inc. **AW** is employed by Parexel.

*Corresponding Author:*
Brian D. Williamson
Biostatistics Division
Kaiser Permanente Washington Health Research Institute
Seattle, WA 98101
(206) 287-2024
brian.d.williamson@kp.org

*Keywords:* real-world data, real-world evidence, causal inference, safety surveillance, reproducibility

**ABSTRACT**

Real-world data, such as administrative claims and electronic health records, are increasingly used for safety monitoring and to help guide regulatory decision-making. In these settings, it is important to document analytic decisions transparently and objectively to ensure that analyses meet their intended goals. The Causal Roadmap is an established framework that can guide and document analytic decisions through each step of the analytic pipeline, which will help investigators generate high-quality real-world evidence. In this paper, we illustrate the utility of the Causal Roadmap using two case studies previously led by workgroups sponsored by the Sentinel Initiative—a program for actively monitoring the safety of regulated medical products. Each case example focuses on different aspects of the analytic pipeline for drug safety monitoring. The first case study shows how the Causal Roadmap encourages transparency, reproducibility, and objective decision-making for causal analyses. The second case study highlights how this framework can guide analytic decisions beyond inference on causal parameters, improving outcome ascertainment in clinical phenotyping. These examples provide a structured framework for implementing the Causal Roadmap in safety surveillance and guide transparent, reproducible, and objective analysis.

**INTRODUCTION**

The Food and Drug Administration's (FDA) Sentinel Initiative is a program for actively monitoring the safety of regulated medical products.[1,2] The Sentinel Initiative uses routinely-collected healthcare databases generated from insurance claims and electronic health records (EHRs) to supplement randomized clinical trials to provide evidence on the real-world effectiveness and safety of pharmaceutical drugs and other

FDA approved products. However, extracting valid evidence from these data sources to help guide regulatory decisions remains challenging due to bias in causal estimates stemming from non-randomized treatments and poorly measured information on patient characteristics and clinical features.

In this paper, we illustrate the application of the Causal Roadmap in two case studies that were previously led by workgroups sponsored by the Sentinel Initiative. Each case example focuses on different aspects of the analytic pipeline for drug safety monitoring. In the first case study, we show how the Causal Roadmap can be used to promote transparency, reproducibility, and objective decision making for causal analyses. In the second case study, we illustrate how the principles of the Causal Roadmap extend beyond causal parameters and can be used to guide analytic decisions for clinical phenotyping for improved outcome assessment.

Each case study was previously conducted and further details can be found on the FDA's Sentinel webpage.[3] It is important to emphasize that the purpose here is not to provide a thorough overview of all decisions made throughout the analytic process for each study, or argue that all analytic decisions within each example are optimal. Instead, our goal is simply to give a high-level overview of how the Causal Roadmap could be applied to settings similar to the previously conducted case examples described here.

**Case Study 1: Enhancing Causal Inference in the Sentinel System: Targeted Learning for Large-Scale Covariate Adjustment in Healthcare Database Studies**

Confounding remains a primary challenge in real-world evidence (RWE) studies for drug safety monitoring. During the early periods of post-market approval, some important confounding factors are often unknown to investigators or not directly measured in these data sources. To improve confounding control in these settings, data-driven algorithms can be used to supplement investigator-specified variables by leveraging the large volume of information in these data sources to generate and identify features that indirectly capture information on unmeasured factors ('proxy confounders').[4,5] However, there continues to be an increasing number of methods available for large-scale causal inference with many variations in how these tools can be applied for high-dimensional proxy confounder adjustment in healthcare databases. Various methods rely on different assumptions that hold in different cases, and it is unlikely that any single approach is optimal across all databases and research questions. Consequently, a fundamental challenge when estimating causal effects in healthcare databases is making objective decisions between alternative analytic approaches while tailoring analyses to the study at hand.

In this case study, we applied the Causal Roadmap to an observational study to assess the impact of nonselective nonsteroidal anti-inflammatories (NSAIDs) vs opioid use on acute kidney injury (AKI). We show how the Causal Roadmap can help to improve transparency, reproducibility, and objective decision making across all aspects of the analytic pipeline. We give particular focus on illustrating the use of outcome-blind simulations to maintain objectivity during the model selection process to tailor analytic decisions for data-driven large-scale covariate adjustment to the given study.

***Overview of study objectives.*** We are interested in understanding the short-term (6-month) impact of initiating NSAID treatment on AKI relative to initiating treatment with an opioid in patients diagnosed with osteoarthritis. NSAIDs and opioids are among the most commonly used pharmacotherapies for pain in patients with osteoarthritis and the safety of these alternative analgesics on AKI is unclear.[6] Confounding was the primary concern in this study and we were interested in applying methods for large-scale covariate adjustment to evaluate if these tools are likely to improve confounding control when estimating the causal effect in this setting.

Consistent with the ICH E9(R1) attributes of a statistical estimand, we define the following:

- *population of interest:* Medicare beneficiaries linked to EHR data from the Research Patient Data Registry (RPDR) at Mass General Brigham (the largest healthcare provider in Massachusetts). The study population was restricted to patients who had continuous enrollment in Medicare parts A, B, and D in the 365 days prior to treatment initiation and were diagnosed with osteoarthritis (defined as having a diagnostic code for osteoarthritis in the 365 days prior to treatment initiation).
- *study treatment:* The treatment was defined as filling at least one prescription for an NSAID after a 365-day washout period of no use (no prescription fill) for any opioid or NSAID (new-user design).[7,8] Similarly, the comparator group was defined as filling a prescription for an opioid after a 365-day washout period having no prescription fill for any opioid or NSAID.

- *outcome*: The outcome was defined as any diagnosis for AKI within 6 months of follow-up. The outcome was identified using a previously developed algorithm for identifying AKI in administrative claims data.[9]
- *summary measure*: 6-month risk difference

**Specify Causal Model and Causal Parameter**

In this example, we chose to model the effect of initiating the treatment vs comparator on the 6-month risk of AKI ('point-treatment' effect).[10] We chose to target the point-treatment effect because the assumptions for identification for this causal parameter are less strict compared to an 'as-treated' analysis where individuals are censored at treatment discontinuation/switching. In administrative healthcare databases, reasons for discontinuation and switching are often not well measured, making identification of an as-treated causal parameter more challenging.

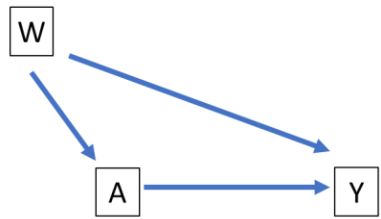

*Figure 1. Directed acyclic graph (DAG) showing a simplified causal model for the case example comparing the effect of initiating NSAID vs opioid on AKI. In this causal model, we make the simplifying assumption that the only bias for the effect of treatment (A) on the outcome (Y), is confounding by W (representing a high-dimensional set of confounders and/or proxies for confounders).*

The causal model for a point treatment effect is presented in the directed acyclic graph (DAG) shown in Figure 1. It is important to note that the causal model is based on underlying causal assumptions which cannot be empirically verified (and therefore requires subjective decisions by investigators). This case study is focused primarily on evaluating different methods of confounder adjustment. The simplified causal model in Figure 1 also emphasizes the possibility of confounding over other issues; for example, in the above DAG, we assume that there is no selection bias or misclassification. Therefore, this DAG does not include censoring events for loss-to-follow-up (death and disenrollment). We assume in our causal model that these censoring events are negligible due to the short-term follow-up (discussed further below).

It is important to emphasize that investigators can disagree on the merits of the chosen causal model (DAG). The purpose of the Causal Roadmap is to help investigators be explicit and transparent in the assumptions made by the causal model so that investigators can think carefully about the plausibility of those assumptions and the conditions necessary for identification of the causal parameter defined in terms of the chosen causal model.

Once the causal model is specified, the causal parameter can then be defined. In this example, we defined the causal parameter as the 6-month risk difference for the outcome. More formally, we define $Y^1$ as the potential outcome under the study drug (NSAID), and $Y^0$ as the potential outcome under comparator drug (opioid). Our causal parameter is given by $\psi^{causal} = E(Y^1) - E(Y^0)$.

**Define observed data and decisions for cohort construction.**

Our dataset consists of linked Medicare-EHR data between the years 2007-2017. We followed the 'Target Trial' framework when constructing our cohort to emulate components of a randomized trial using observational data.[11,12] Because treatment is not randomized, we defined treatment groups using the 'new-user' design as discussed previously. This design for constructing treatment groups helps avoid bias by having a clear time-frame for treatment initiation and follow-up as in randomized clinical trials. By having a clear time-frame for the start of follow-up, the new-user design helps to avoid biases caused by conditioning on variables on the causal pathway and avoids comparing current (prevalent) users to new-users of the treatment and comparator groups of interest.[8]

After identifying new-users of the treatment and comparator groups, we restricted the cohort to individuals diagnosed with osteoarthritis. This was defined as having a diagnostic code for osteoarthritis in the 365 days prior to treatment initiation. After restricting to individuals with osteoarthritis, the new-user cohort included 21,343 individuals with 7,767 (36.4%) individuals initiating NSAIDs, 13,576 (63.6%) individuals initiating Opioids, and 899 (4.2%) individuals having an outcome event. Baseline covariates available for adjustment consisted of 91 investigator-specified variables and an additional 14,938 features available for proxy adjustment. These additional proxy features were derived from all claims codes and codes from EHR structured data after screening codes with a prevalence <0.001.

**Statistical estimand and Identifying assumptions**

We defined the statistical target of estimation as the marginal risk difference, $\psi^{obs} = E[E(\Delta Y \mid \Delta = 1, A = 1, W)] - E[E(\Delta Y \mid \Delta = 1, A = 0, W)]$, where $\Delta = 0$ indicates the outcome was censored and Δ= 1 indicates it was observed in the data. $\psi^{obs}$ is equivalent to $\psi^{causal}$ under the following set of identifying assumptions: 1) consistency, 2) conditional exchangeability (no unmeasured confounding or selection bias), 3) positivity.

If we are willing to make an additional assumption of uninformative right censoring (MCAR), we can define the statistical estimand as $\psi'^{obs} = E[E(Y \mid A = 1, W) \mid \Delta = 1] - E[E(Y \mid A = 0, W) \mid \Delta = 1]$, asserting that the statistical parameter in the sub-population where the outcome is observed is equivalent to the statistical parameter in the full study population.

Before proceeding with estimation of the target parameter, it is important to consider the plausibility of each assumption necessary for identification of the causal parameter. Diagnosis codes for AKI are known to be highly sensitive and specific, so the consistency assumption is likely to be satisfied. It is possible (and even likely) that conditional exchangeability is not fully satisfied, because some degree of baseline confounding is likely due to unobserved factors and there may be some selection bias due to informative censoring. However, given the short follow-up, less than 4% of study participants were censored due to death or disenrollment, and censoring was similar in each study arm. To simplify analyses for this case study focused on confounder adjustment, and because reasons for censoring are often not captured well in administrative healthcare data, we were willing to make the assumption of uninformative right censoring.

**Estimation**

Estimation of $\psi'^{obs}$ was carried out on the dataset omitting observations where outcomes were censored due to loss-to-follow-up (<4% of observations). As discussed previously, while this can induce a selection bias, we were willing to assume that the impact was negligible due to the low degree of censoring. We used targeted minimum loss-based estimation (TMLE) to estimate the average treatment effect in the population. For the outcome model used within TMLE, we fit a Lasso regression to optimize cross-validated prediction. However, in this study, outcome events were rare and it was difficult to fit large-scale models for the outcome. Therefore, for all TMLE models, we focused on large-scale covariate adjustment through modeling the treatment assignment (the propensity score [PS]).

When modeling the PS within the TMLE framework, we compared 8 Lasso-based PS models for large-scale covariate adjustment. We briefly outline PS Models 1 through 8 below:

- *Model 1: Traditional Lasso*: Logistic regression model using L1 regularization (Lasso), where the loss function for choosing the lambda tuning parameter (degree of regularization) is based on minimizing the out-of-sample (cross-validated) predictive performance for treatment.[13]
- *Model 2: Outcome Adaptive Lasso*: We applied a variation of the outcome adaptive lasso proposed by Shortreed & Ertefaie.[14] Our variation consisted of first fitting a lasso model for the outcome and identifying all variables whose coefficient was not shrunk to zero. We then fit an Adaptive Lasso model for treatment assignment to allow for specification of different penalization weights for different variables. All variables whose coefficient in the outcome lasso model was not shrunk to zero were forced into the treatment Lasso model by assigning them a penalization weight of zero. All other variables were penalized similarly to Model 1 (based on optimizing cross-validated predictive performance for treatment).

- *Model 3: Collaborative-Controlled Lasso*: The collaborative-controlled lasso is a recently proposed extension of Lasso regression for purposes of estimating the PS.[15] Instead of choosing the lambda tuning parameter to minimize cross-validated prediction for treatment assignment, the collaborative-controlled lasso uses the principles of collaborative targeted learning to consider a bias-variance tradeoff in the estimated treatment effect when selecting the degree of regularization (lambda tuning parameter).[15]

- *Model 4: Collaborative-Controlled Outcome-Adaptive Lasso*: This approach combines Collaborative Learning with Model 2 described previously. The first step in Model 4 is the same as Model 2 (variables selected by the outcome Lasso are forced into the treatment model). However, when fitting the treatment adaptive lasso model in the second step, Model 4 uses collaborative targeted learning to select the regularization tuning parameter instead of using cross-validated prediction for treatment.

- *Models 5 through 8*: Models 5 through 8 are equivalent to Models 1 through 4, except that they incorporate cross-fitting when modeling treatment assignment and assigning predicted values for the propensity score. Cross-fitting (sample splitting) has been recommended when using data-adaptive (machine learning) algorithms for estimating nuisance models for causal inference (e.g., the PS and outcome model).[16-18] Here, we only considered application of cross-fitting to the PS model to reduce problems of nonoverlap caused by modeling spurious associations in the PS.

*Outcome-blind simulations*: In order to choose between the alternative models described above, we need an objective framework. The use of synthetically generated datasets that combine real data from the given study with simulated causal effects has become increasingly popular to help tailor analytic choices for causal inference.[19-23] Frameworks for generating synthetic datasets have largely been based on use of the parametric bootstrap.[19] Here, we applied a similar simulation approach to provide objective empirical guidance for model selection. Briefly, we bootstrapped subjects from the observed data structure and left associations between baseline covariates unchanged. We then injected causal relations between a subset of variables to simulate treatment assignment as well as the outcome. This allowed us to generate data with a known treatment effect while maintaining some of the complexity of the observed data structure to compare statistical properties of different analytic approaches. The goal of these outcome-blind simulations is to help investigators tailor analytic decisions to the given study while maintaining objectivity during the analytic process.

**Study Results**

*Results for Outcome-Blind Simulations*: In Figure 2 we present outcome-blind simulation results for the empirical study. Figure 2 shows that the collaborative controlled extensions of the Lasso and Outcome-Adaptive Lasso when using cross-fitting (Models

7 and 8) had similar performance, with both these approaches outperforming the other Lasso models in terms of bias (Plot A), MSE (Plot B), and coverage (Plot C). Overall, the collaborative controlled outcome adaptive lasso with cross-fitting (Model 8) performed best when implemented using TMLE, with a slight incremental benefit over Model 7.

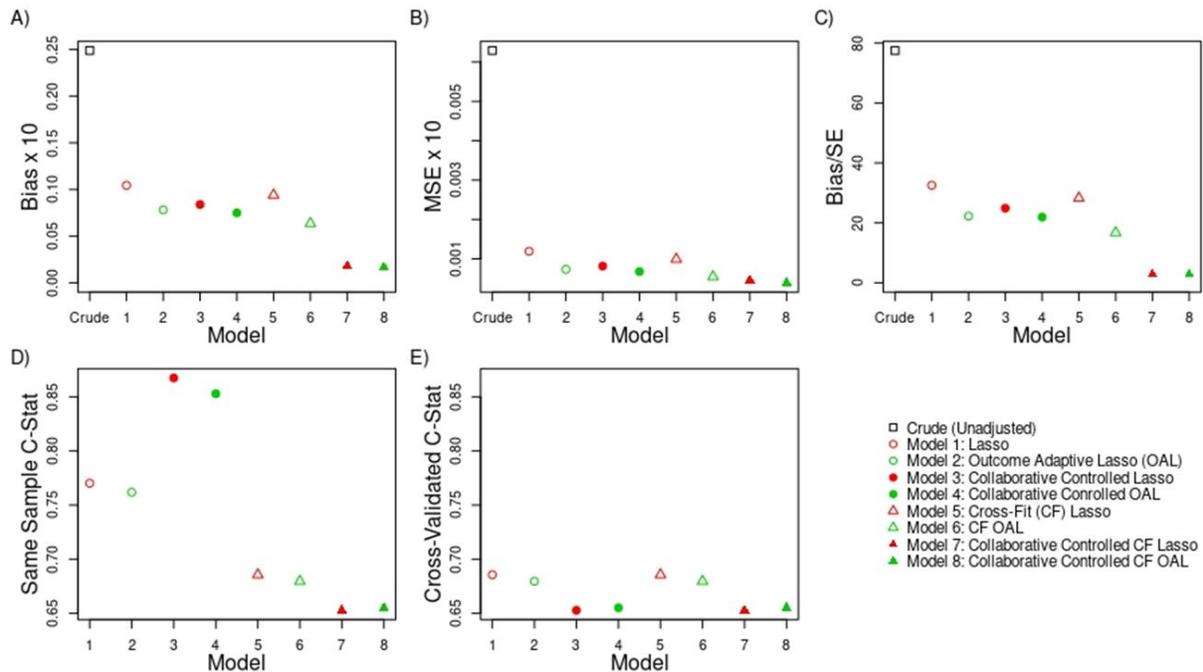

*Figure 2.* Outcome-blind simulation results for study 1.

*Empirical Study Results*: For the outcome-blind simulation study, Model 8 performed best in terms of reducing bias in the estimated treatment effect. Therefore, we applied this model to the empirical study for large-scale covariate adjustment. The unadjusted 6-month risk difference comparing NSAIDS vs opioids on acute kidney injury was 0.024 (0.018, 0.030). After large-scale covariate adjustment using TMLE with a PS that was estimated using the collaborative-controlled outcome-adaptive Lasso with cross-fitting (Model 8) the risk difference was 0.005 (-0.027, 0.038).

**Sensitivity Analysis**

Causal bias is the gap between the statistical and causal parameter that can arise when any causal assumptions are violated. A non-parametric sensitivity analysis illustrates how departures from causal assumptions would impact study findings.[24] The sensitivity plot shows the shift in point estimates and 95% confidence interval bounds under a range of gap sizes (Figure 3). Gap size can equivalently be expressed in units that facilitate a basis for comparison, such as effect size units ($\delta$) or relative to the bias adjustment due to measured confounders (Adj units). Bounds on the plausible gap size can be obtained from external knowledge and through the use of negative controls. The G-value is the size of the minimal causal gap that would negate the study finding.[25] In our example the G-value = 0.027 on the risk difference scale.

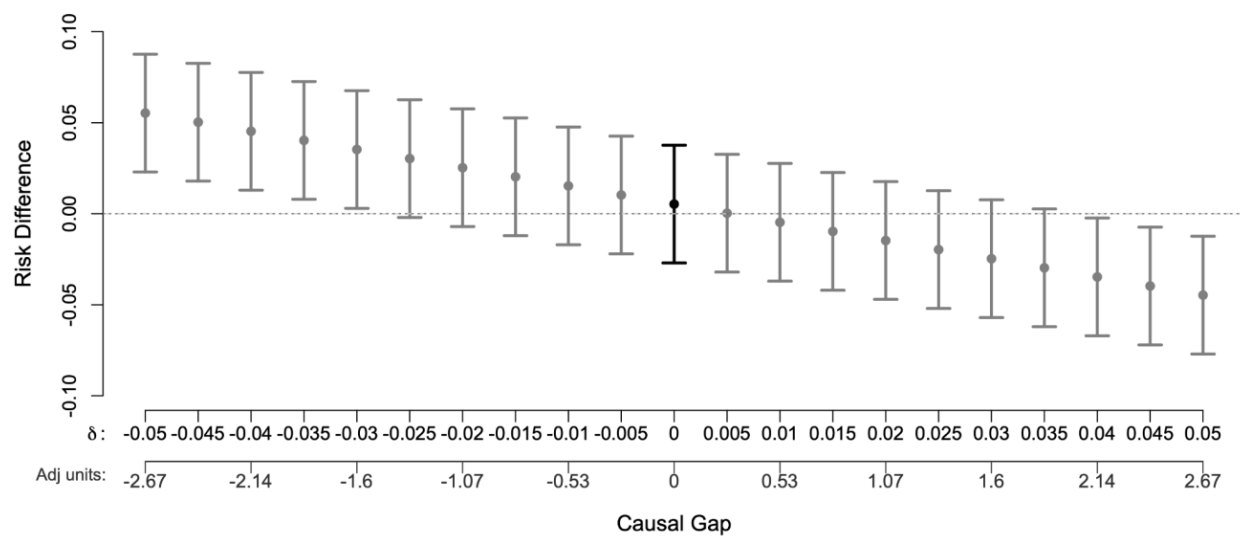

*Figure 3.* Estimated effects and 95% confidence intervals from a sensitivity analysis (intervals in gray) showing how departures from the assumption of exchangeability (no unmeasured confounding, informative censoring, etc.) would impact the calculated risk difference and 95% confidence interval (interval in black).

**Case Study 2: Scalable Prediction of Safety Outcomes Using Electronic Health Record Data**

Safety studies are run within Sentinel only when the ARIA (Active Risk Identification and Analysis) system is deemed sufficient to answer the safety question at hand, based on the available data and methods.[26,27] The Sentinel distributed database primarily includes claims data from 17 Sentinel data partners;[28,29] some of these data partners also provide EHR data. ARIA sufficiency judgments rely on understanding whether exposure to the study drug, comparator, and health outcome of interest (HOI) can be accurately assessed from these observational data. If ARIA is deemed insufficient, then FDA may require the sponsor to run a postmarketing study.

The end goal of a Sentinel safety surveillance study is often to determine if a medical product is causing unintended adverse effects. However, for some health outcomes of interest it may be difficult to identify the health outcome of interest from information in claims data due to a lack of key information. In the remainder of this section, we discuss how principles underlying the Causal Roadmap can guide the use of machine learning to identify an outcome from data – a classification activity known as "phenotyping" in the clinical informatics literature.[30]

A distinction must be made about the phenotyping process (which is fundamentally a prediction problem) and the downstream use of the phenotype (e.g., estimating prevalence, to satisfy cohort inclusion criteria, outcome in a causal inference problem). Here we focus on identifying case status to be used as a binary outcome in a downstream retrospective safety study. Publications have shown that non-differential misclassification of the outcome biases the estimate of an additive treatment effect, but not a relative risk (RR).[31] However, using a classifier that has a low positive predictive value (PPV) will bias a RR estimate, while using a classifier that has a low sensitivity will increase the variance of the RR estimate. Thus, high PPV and high sensitivity are desirable properties. Although phenotyping is a predictive modeling task rather than a causal inference task, a variant of the Causal Roadmap can serve as a guide.

We now discuss several steps from the Causal Roadmap that can serve as a guide throughout a prediction modeling task. Our example phenotyping task is to predict anaphylaxis using structured medical claims and EHR data (both structured data and unstructured text, with natural language processing methods used to extract information from the unstructured text); the results of this analysis were published by Carrell et al.,[32] who followed the steps that we highlight below.

**Step 1a: Specify the study question**

Anaphylaxis is a rare, though life-threatening, systemic allergic reaction that occurs shortly (minutes to hours) after contact with an allergy-causing substance (e.g., food, medication, or insect bite).[33] Anaphylaxis diagnosis codes are a poor proxy for true anaphylaxis events. Approximately 1/3 of medical encounters for which an anaphylaxis code exists are not true cases of anaphylaxis.[34] Carrell et al.[32] developed a phenotyping algorithm using claims data augmented with labs and EHR data with the goal of improving the positive predictive value (PPV) for anaphylaxis without sacrificing sensitivity.

For our classification task, we will define the phenotype as an indicator Y of the predicted probability of having an anaphylaxis event above a threshold. In terms of the full counterfactual data the parameter of interest for the prediction task is $\psi^* = E(Y \mid X)$, where $E$ denotes the expectation and $X$ contains measures of the relevant features guided by a downstream use of the phenotype as the outcome in a future safety study.

**Step 1b: Specify the causal model**

Although phenotyping is a prediction task rather than a causal one, we will keep its potential downstream use as an outcome in a retrospective study in mind when choosing features to consider as candidate predictors. Consider a generic directed acyclic graph (DAG) for such a study, where A is an indicator of treatment (Figure 4).

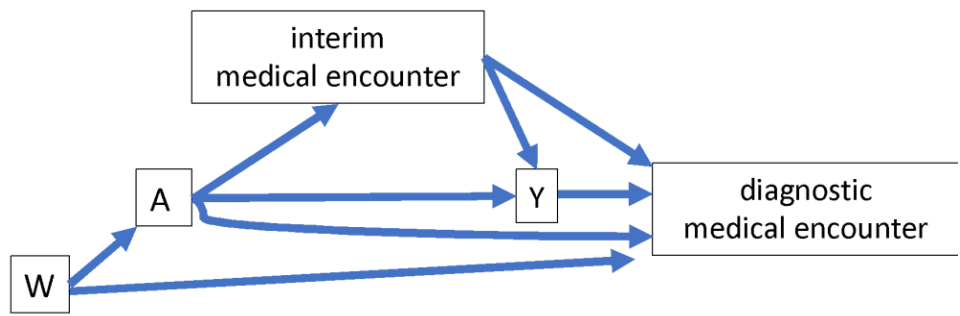

*Figure 4: an example directed acyclic graph (DAG) showing a possible causal model relating baseline covariates, W; an exposure of interest, A; an outcome of interest, Y; and variables captured during interim or diagnostic medical encounters.*

Baseline covariates, $W$, are ideal candidate features to include in the prediction model. Claims and EHR data connected to the diagnostic medical encounter are central to recognizing the occurrence of the outcome event. Each medical encounter offers the opportunity to capture information on diagnosis and prescription codes, symptoms, lab orders, lab values, and time-varying patient characteristics. The DAG in Figure 4 illustrates that if exposure (A) leads to differential capture of these covariates, then their inclusion in a predictive model could bias the downstream effect estimate. This may affect the interpretation of the downstream causal effect estimate by conflating the effect of pre-planned monitoring with the effect of the study drug. These issues are unlikely to arise in our study since anaphylaxis typically occurs within minutes or hours after exposure.

**Step 2: Define observed data**

The study population is defined as Kaiser Permanente Washington (KPWA) members who had a qualifying health encounter between October 1, 2015 and March 31, 2019. KPWA is an integrated-care delivery organization operating in Washington State and northern Idaho. Qualifying health encounters consisted of (a) an emergency department or inpatient encounter with an anaphylaxis diagnosis, and (b) an outpatient encounter with an anaphylaxis diagnosis and (on the same day from any setting) either a diagnosis code for one of several conditions that often co-occur with anaphylaxis or a procedure code for one of several procedures that are often used to treat anaphylaxis.[32,35]

Clinician resources limited the number of charts reviewed to ascertain gold standard outcomes. The analytic dataset consisted of 239 people. Medical records for each person's qualifying encounter were reviewed to determine whether potential events met the National Institute for Allergy and Infectious Disease clinical criteria for anaphylaxis. Of the 239 people, 154 were found to have a validated anaphylaxis event (Y = 1), while 85 were found to not have a validated anaphylaxis event (Y = 0). The covariates, X, include 159 potential predictors, including 43 structured covariates (e.g., demographics, potential cause of anaphylaxis, and history of allergic reactions obtained from the Sentinel distributed database) and 116 natural language processing (NLP)-derived covariates from the KPWA EHR. Domain knowledge informed the selection of features likely to discriminate between cases and non-cases within the population satisfying the inclusion criteria.

**Step 3: Assess identifiability**

In this case study, the purpose is to predict an observed outcome given covariates, rather than to make causal inference. We do not consider a counterfactual outcome – in other words, the observed data are identical to the ideal data necessary to answer the question of interest. While this step is not relevant to the immediate purpose, it is relevant for downstream causal analyses using predicted outcomes. For example, maximizing PPV may reduce bias due to misclassification, while maximizing sensitivity may increase power to detect an effect.

**Step 4: Define the statistical estimand**

The statistical estimand is $\psi^* = E(Y \mid X)$.

**Step 5: Choose a statistical model and estimator**

Any estimator of a prediction function (equivalent to a conditional mean function, our statistical estimand) is constrained by the amount of information in the data, which is governed by the sample size (for continuous outcomes) or effective sample size (the number of observations in the minority class, for binary outcomes).[36] To avoid overfitting to the data, Carrell et al.[32] first applied outcome-blind dimension reduction techniques that reduced the feature set to $Z$, consisting of 132 covariates. Prediction models were fit using structured features only and structured and NLP-derived features, to determine if the NLP-derived features conferred a benefit to anaphylaxis identification.

Rather than focusing on a single regression technique, Carrell et al.[32] evaluated 25 parametric and machine learning algorithms, using cross-validation to estimate the cross-validated area under the receiver operating curve (cvAUC). The first 24 consisted of all combinations of eight base learners (logistic regression, elastic net, two variants of gradient boosted trees, two variants of Bayesian additive regression trees (BART), and two neural network architectures), each coupled with three pre-screening algorithms (retain all, partitioning around medoids, and Lasso). The final algorithm was the Super Learner, an ensemble method defined as an optimal weighted combination of the individual candidate algorithms.[37]

For each prediction function, Carrell et al.[32] estimated both cvAUC and cross-validated estimates of classification performance metrics at candidate classification thresholds to illuminate the tradeoffs between PPV and sensitivity.

**Results**

Carrell et al.[32] used the Causal Roadmap principles throughout their analysis, which enabled a clear comparison of results across a variety of prediction techniques and input feature sets. They showed that the use of machine learning far exceeded logistic regression in predicting anaphylaxis in high-dimensional data, and combining NLP-derived features with structured data conferred additional benefits. The cvAUC for a main terms logistic regression model using structured data only was 0.58, while the prediction function produced using machine learning trained on all available features was 0.70.

The authors also evaluated cvAUC for each of the models developed using the KPWA data on data from Kaiser Permanente Northwest (KPNW), an integrated-care delivery organization in northwest Oregon and southwest Washington State, collected using identical methods to the KPWA data. They found that there was a modest degradation in prediction performance in this new population.

The maximum PPV observed at KPWA was 86%. The cutpoint of predicted risk yielding a PPV of 78.7% at KPWA yielded a sensitivity of 65.8%; this same cutpoint yielded a 78.1% PPV and 55.6% sensitivity at KPNW. Thus, in a downstream causal analysis there will be outcome misclassification. This is a violation of the consistency assumption that should be considered in any sensitivity analysis, e.g. the PPV suggests bias in an estimated relative risk due to this violation will be mild. However, the relatively low sensitivity on the external validation data indicates decreased power to detect a statistically significant effect.

**DISCUSSION**

Case Study 1 illustrates how the roadmap guides breaking down a complex problem into manageable sub-parts that clearly describe the chain of reasoning from study question to study finding. This case study further highlights how the roadmap can help facilitate transparent and objective analytic decisions throughout the analytic pipeline, with a specific focus on using outcome-blind simulations to tailor analytic choices for large-scale covariate adjustment. In post-market safety analyses, it is often difficult to know the optimal analytic approach for covariate adjustment. When the optimal analytic approach is not known, outcome-blind simulations allow investigators to tailor analytic choices to the study at hand, while maintaining objectivity during the analytic process by not letting information on the treatment-outcome association contribute to decisions on model selection.

In this case study, the outcome blind simulations provided evidence that the use of collaborative learning when fitting large-scale PS models can help reduce bias in the estimated treatment effect. After applying the selected approach to the empirical example, we found that there was no evidence for increased risk of AKI between patients taking opioids vs NSAIDs. While hidden biases—including unmeasured confounding, selection bias, and misclassification—could impact our results, the sensitivity analysis suggests that bias would have to be large to explain away the observed null effect.

Case study 2 illustrates how to apply the Causal Roadmap and rigorous thinking to a predictive modeling project, and a framework for making performance tradeoffs relative to a final goal (e.g., estimating a causal effect in a safety study). The Causal Roadmap encourages documenting the decisions and choices made during the course of the analysis, and how each can impact the statistical result, as well as the estimation of the true target parameter (both in the prediction task in the case study and the eventual safety analysis).

This case study also highlights the role that outcome misclassification can play in the estimation of causal effects. While there were no identification assumptions necessary to estimate the conditional mean outcome given covariates (our target in the prediction task), the consistency assumption (that an individual's potential outcome given their exposure history is equal to the observed outcome), which is crucial for causal inference on, e.g., a causal relative risk, may be violated if the outcome is misclassified.[38,39] Attention to the consistency assumption can help determine the prediction performance metrics to focus on when determining if a phenotyping algorithm (or more generally, an outcome identification procedure) achieves satisfactory performance to use in the downstream safety study.

**CONCLUSION**

The Causal Roadmap is a useful tool to guide analytic decisions in safety studies (including for both causal inference and prediction). The roadmap guides breaking down a complex problem into manageable sub-parts that clearly describe 1) the chain of reasoning from study question to study finding; 2) which aspects of the analysis involve observable data and which require additional assumptions; and 3) sensitivity analyses to assess the validity of interpretation as a causal effect. Many of the ideas that make up the Causal Roadmap are not new – it ties together a rich literature and history of making principled decisions in biomedical and public health research – but it provides a structured and reproducible approach. Finally, real-world data (including claims and EHR) is becoming increasingly used in safety studies to help guide regulatory decision making. In this context, the Causal Roadmap can help facilitate transparent documentation and objective analytic decisions to help researchers better understand how well analyses using real-world data meet the intended goals.